\documentclass[sigconf,authorversion,nonacm]{acmart}
\usepackage[utf8]{inputenc} 
\usepackage[T1]{fontenc}    
\usepackage{booktabs}       
\usepackage{amsfonts}       
\usepackage{amsmath}        
\usepackage{nicefrac}       
\usepackage{microtype}      
\usepackage{lipsum}		    
\usepackage{graphicx}
\usepackage{xcolor}         
\usepackage{setspace}       
\usepackage{subcaption}     
\usepackage{verbatim}       
\usepackage{placeins}       
\usepackage{hyperref}
\usepackage{comment}

\newcommand\numberthis{\addtocounter{equation}{1}\tag{\theequation}}

\AtBeginDocument{%
  \providecommand\BibTeX{{%
    \normalfont B\kern-0.5em{\scshape i\kern-0.25em b}\kern-0.8em\TeX}}}

\begin{document}

\title{Regime-based Implied Stochastic Volatility Model for Crypto Option Pricing}
\author{Danial Saef}
\authornote{Support from the German Research Foundation [IRTG 1792]; the European Union’s Horizon 2020 “FIN-TECH” Project  [825215, Topic ICT-35-2018, Type of actions: CSA]; the European Cooperation in Science \& Technology COST Action [CA19130]; the Czech Science Foundation [19-28231X, CAS: XDA 23020303]; and the Yushan Scholar Program of Taiwan greatly acknowledged.}
\orcid{}
\authornotemark[1]
\affiliation{%
  \institution{Humboldt University of Berlin}
  \streetaddress{Unter den Linden 6}
  \city{Berlin}
  \state{}
  \country{Germany}
  \postcode{10178 }
}
\email{danial.saef@hu-berlin.de}

\author{Yuanrong Wang}
\affiliation{%
  \institution{University College London}
  \streetaddress{66-72 Gower Street}
  \city{London}
  \country{UK}
  \postcode{WC1E 6BT}}
\email{yuanrong.wang@cs.ucl.ac.uk}

\author{Tomaso Aste}
\affiliation{
  \institution{University College London}
  \streetaddress{66-72 Gower Street}
  \city{London}
  \country{UK}}
\email{t.aste@ucl.ac.uk}

\renewcommand{\shortauthors}{Saef, Wang and Aste}

\begin{abstract}
\noindent The increasing adoption of Digital Assets (DAs), such as Bitcoin (BTC), raises the need for accurate option pricing models. Yet, existing methodologies fail to cope with the volatile nature of the emerging DAs. Many models have been proposed to address the unorthodox market dynamics and frequent disruptions in the microstructure caused by the non-stationarity, and peculiar statistics, in DA markets. However, they are either prone to the curse of dimensionality, as additional complexity is required to employ traditional theories, or they overfit historical patterns that may never repeat. Instead, we leverage recent advances in market regime (MR) clustering with the Implied Stochastic Volatility Model (ISVM). Time-regime clustering is a temporal clustering method, that clusters the historic evolution of a market into different volatility periods accounting for non-stationarity. ISVM can incorporate investor expectations in each of the sentiment-driven periods by using implied volatility (IV) data. In this paper, we apply this integrated time-regime clustering and ISVM method (termed MR-ISVM) to high-frequency data on BTC options at the popular trading platform Deribit. We demonstrate that MR-ISVM contributes to overcome the burden of complex adaption to jumps in higher order characteristics of option pricing models. This allows us to price the market based on the expectations of its participants in an adaptive fashion.
\end{abstract}


\maketitle
\section{Introduction}
Digital Assets have gained massive attention, but research on option pricing in the field is only slowly gaining traction. Their highly volatile nature raises the need for a better understanding of their statistical properties. This is necessary for fitting meaningful models e.g. for option pricing. Most commonly, such models are fitted based on historic return data. However, this approach requires that future movements can be explained at least partially by past patterns, which is questionable for the new, highly speculative class of DAs. Moving away from the reliance on historic patterns, Aït-Sahalia et al. \cite{ait-sahalia_implied_2021} show how the implied volatility (IV) surface can be used to fit Implied Stochastic Volatility (ISV) Models. Using IV allows us to fit option pricing models based on market expectations, which gives us the unique opportunity to incorporate the aggregated knowledge of informed traders. Alexander et al. \citep{alexander_net_2022} find that buying pressure in DA option markets is largely driven by informed traders. Their buying pressure, along with speculation and sentiment causes frequent jumps as pointed out by Saef \cite{saef_understanding_2021}. While including jumps in ISV models is possible, such jumps are limited to the return dynamics, and accounting for other factors would require substantially altering the procedure and extending the theory behind them. The current procedure requires extremely high frequency data that is impossible to obtain due to liquidity constraints in DA option markets. Additionally, jumps in DA are still not well defined due to the unique market microstructure of DA markets. Since jumps change the properties of the observed time series it is crucial to account for them. To overcome these empirical and theoretical challenges, we instead propose fitting multiple models under various volatility regimes where the subset of time-series has a stable characteristic. The implied volatility regimes are characterised in a multivariate setup of options with different strike prices and maturities. Our approach extends previous approaches such as Häusler \cite{hausler_cryptocurrency_2021} who have clustered different regimes for Stochastic Volatility (SV) models but in a univariate setup using k-means clustering. 

\begin{figure}[ht]
\centering
\begin{subfigure}{.3\textwidth}
\includegraphics[width=\textwidth]{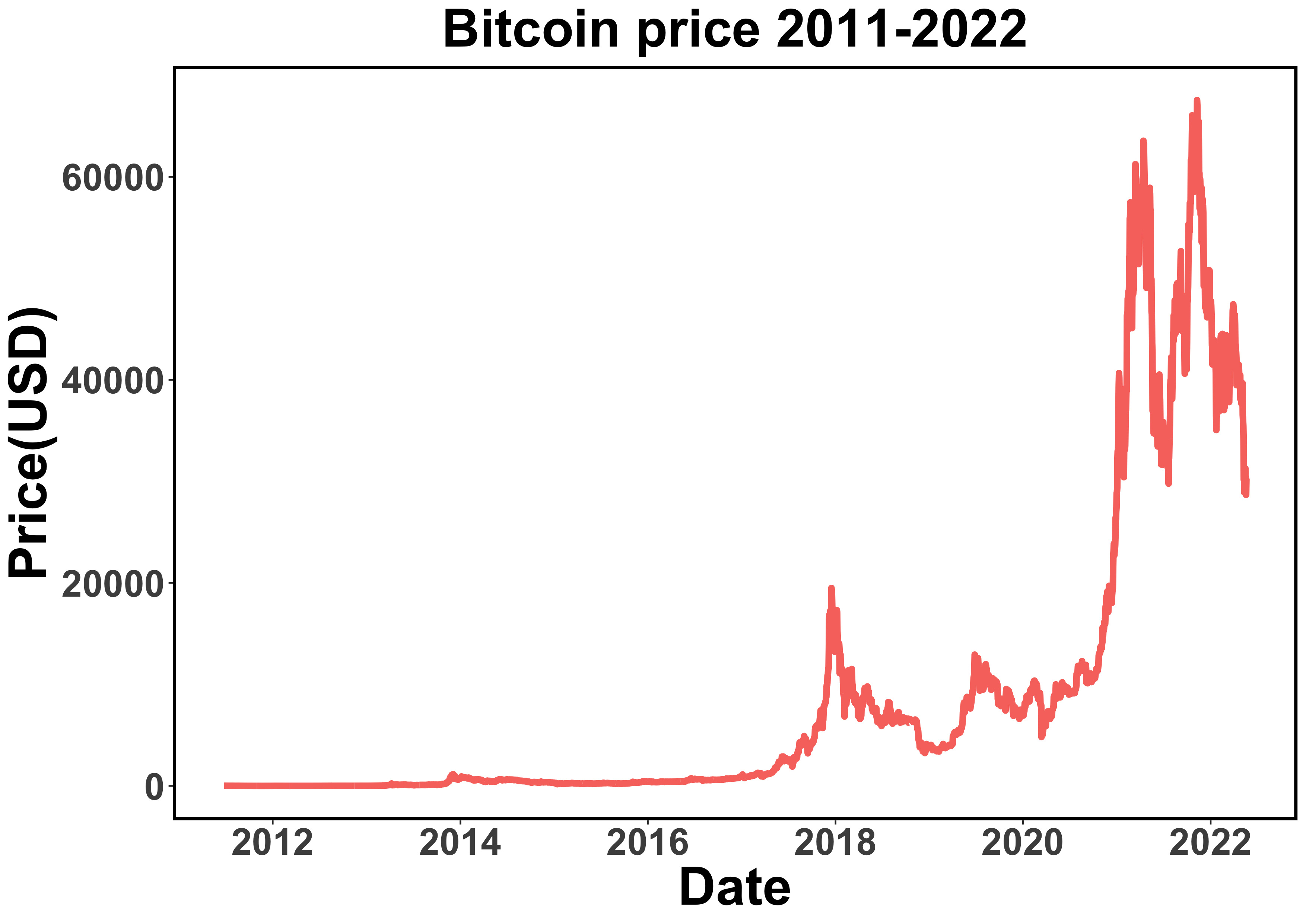} 
\end{subfigure}

\begin{subfigure}{.3\textwidth}
\includegraphics[width=\textwidth]{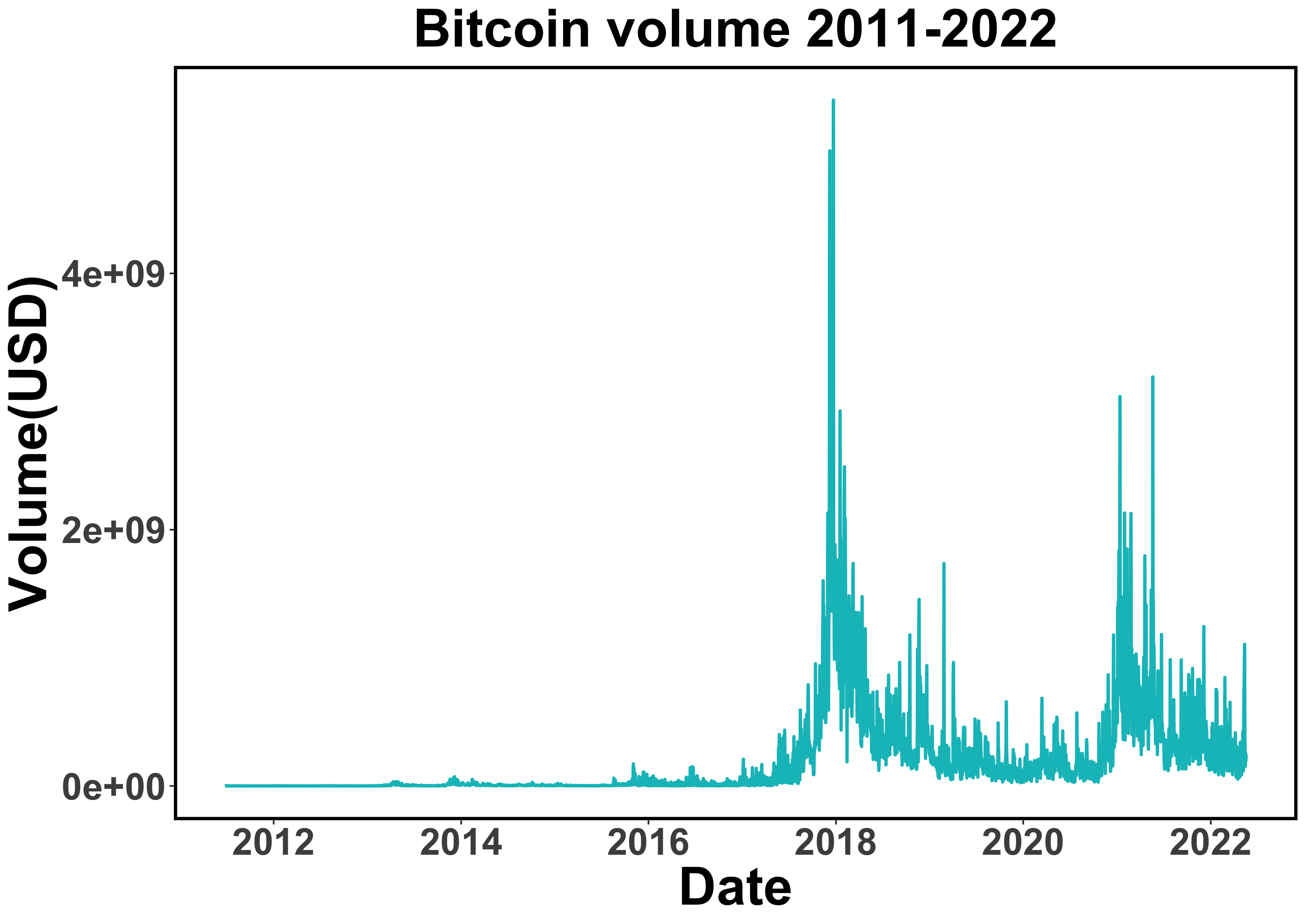}
\end{subfigure}

\begin{subfigure}{.3\textwidth}
\includegraphics[width=\textwidth]{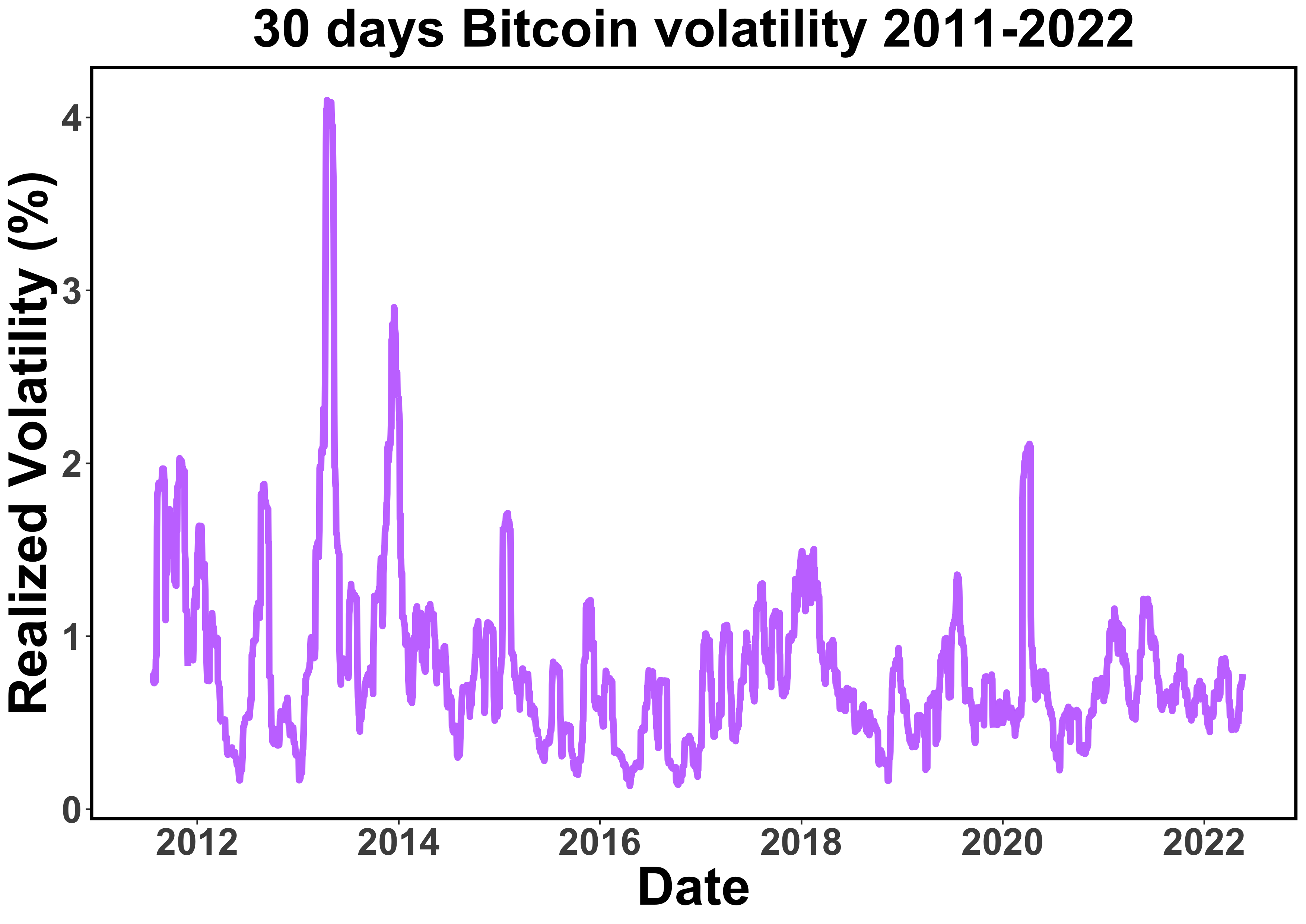}
\end{subfigure}

\caption{Bitcoin price, volume and realized volatility from 2011 to 2022. (Source: Quandl) }
\label{fig:btctimechange}
\end{figure}

Moreover, jumps, no matter if in volatility or price, require modeling their appearance and frequency. While there are models for jumps in traditional assets, these models are difficult to adapt to DAs, because of the different market microstructure. Figure \ref{fig:btctimechange} shows the dynamics of Bitcoin prices, trading volume, and realized volatility from 2011-2022. Price and volume have risen exponentially and are subject to extreme fluctuations as of May 2022, while realized volatility has risen and decreased heavily since 2012. It is moreover remarkable that trading volume has been historically high during the crypto bubble in 2018, that in retrospective looks like only a precursor to the most recent all-time-highs that Bitcoin prices have reached. This very unusual behavior raises two major concerns. First, the post-hoc analytical nature of many financial theories is prone to retardation, especially in the fast-changing DA markets. Undergoing research might still fail to address future market developments, as they might have already changed. Second, it is questionable how much explanatory power historical prices have. Regimes seem to change constantly, and it is in May 2022 impossible to predict where prices are headed. Models like Stochastic Volatility with Correlated Jumps (SVCJ) \citep{duffie_transform_2000} generally work well using e.g. historical return data. However in light of the behavior of BTC the question arises whether historical data is still effective to build data-driven models.

We use Bitcoin option data from Deribit, which is the most popular platform for derivative trading and particularly popular with market makers. We obtain the data from the \protect \includegraphics[height=0.3cm]{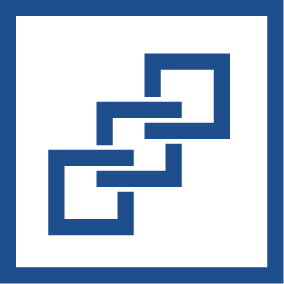}  {\color{blue}\href{https://blockchain-research-center.de}{Blockchain Research Center}} (BRC). We evaluate data from April 13, 2021 - March 7, 2022 and take a rolling window of 5 days. At a frequency of 20 minutes this maintains a balance between having enough data while keeping it recent enough to be meaningful in a high frequency setup.

The main results are that by accounting for the sentiment-driven nature of DAs, we introduce a market regimes-based Implied Stochastic Volatility model, termed MR-ISVM to Bitcoin options in an effort to increase the accuracy of existing option pricing models. This is achieved by using common shape characteristics of the IV surface, more specifically moments of the moneyness and time-to-maturity dimension. By additionally using clustering, we avoid the need for incorporating the unique market microstructure of cryptocurrencies, where jumps in prices or volatility are a common feature. We also avoid making complex model extensions to the overall well working ISVM framework. To the best of our knowledge, this is the first approach of adapting such a model to the digital asset space. The choice of number of clusters and the stability of this selection is well reflected by the fitting of the resulting model. The distribution of the Root Mean Squared Error (RMSE) over all datasets is significantly narrower and lower in mean when using clustering. Finally, the interpretation of clusters is achieved by the comparison of no-clustered, 2-clustered and 3-clustered models with ICC by Proccaci \& Aste \cite{procacci_forecasting_2019}. The interpretation of the cluster outcomes has proven to be difficult when the data is complex or the number of clusters is high. Investigating how an unclustered model behaves against models with multiple clusters thus sheds some light on what these clusters really mean in the context of implied volatility.

\FloatBarrier
\section{Related work} \label{sec:litreview}
This work relates closely to previous research on DAs, picks up on research areas outlined by Härdle \cite{hardle_understanding_2020}, and extends the growing research body on pricing cryptocurrency options. There exists a large branch of literature on option pricing models, Hou et al. have laid a foundation for pricing cryptocurrency options \cite{hou_pricing_2020}. Petukhina studied volatility and trade volume patterns in cryptocurrencies \cite{petukhina_rise_2021}. Scaillet et al. find that Bitcoin is jumping significantly more than traditional assets \cite{scaillet_high-frequency_2020}. Saef et al. show that jumps are a feature of all large cryptos and that option pricing models need to account for them \cite{saef_understanding_2021}. Häusler et al. conclude that crypto option prices are largely driven by jumps in returns and volatility \cite{hausler_cryptocurrency_2021}. However, incorporating jumps into volatility model is non-trivial and convoluted. Instead, by separating market into regimes and fitting volatility model in 

\subsection{Volatility Modeling}

Researchers have shown that DA prices are largely influenced by intra-day jumps caused by news and sentiment instead of historical returns or standard risk factors \citep{liu_risks_2021}. In addition, approaches employing the SVCJ usually focus on daily data \citep{hou_pricing_2020}, such that the behavior of the model in a high frequency (HF) setup is yet to be explored, whereas we argue that the HF setup is especially interesting as DA markets seem to move exponentially faster than traditional markets. The valuation of cryptocurrencies also remains an open question \citep{burda_valuing_2021}, as those assets are driven by sentiment that is difficult to valuate. While there is no unique definition of what comprises sentiment, Baker et al. define it as beliefs about future cash flows and investment risks that are not backed by facts \citep{baker_investor_2007}. In the DA world it is usually represented by observing whether people in online communities such as Twitter or Reddit are majorly bullish or bearish based on specific vocabulary used, see e.g. Nasekin \cite{nasekin_deep_2020}. 

Aït-Sahalia et al. have shown that ISV models can theoretically recover the parameters of models such as SVCJ \cite{ait-sahalia_closed-form_2021}. However, due to empirical challenges it is currently impossible to incorporate a jump component. This reduces researchers to replicate SV type models when using IV data. To overcome this limitation, we propose using a clustering approach to identify different regimes in the data and use it as a pre-filter to fit several models that produce more accurate results than an unfiltered model. The proposed approach works for any number of clusters and without strong assumptions on the number and nature of clusters. The clustering approach follows Procacci \& Aste \cite{procacci_forecasting_2019} and Wang \& Aste \cite{wang_dynamic_2022} closely. This is a time series clustering algorithm based on IV, but with specific modification to serve this project.
Different volatility states exist, and they can be inferred by the associated sparse inverse covariance matrix at each point in time as in Barfuss et al. \cite{barfuss_parsimonious_2016}.

Lai shows that forward-looking information extracted from option-implied equity risk premia can improve regime detection compared to models based on historical return data \cite{lai_detecting_2022}. Using the IV surface to obtain a model instead thus allows us to incorporate investor expectation as a proxy for the current sentiment. We use close to maturity (one week or less) options to cluster the market in an attempt to capture investor expectation. Previous literature suggests that close to maturity options reflect current beliefs about the market most accurately \citep{giglio_excess_2018}. Especially long maturity option prices are unjustifiably variable as investors seem to have a tendency to overemphasize the value of recent data while ignoring other possibly relevant information when establishing their beliefs about the future \citep{stein_overreactions_1989}. 

\subsection{Market Regime Clustering}
Over the past decade, stochastic regime switching models \cite{hamilton_analysis_1990} have proven to be an effective approach for volatility forecasting with regime awareness based on low-frequency financial and economical data. Later, the generalized autoregressive conditionally heteroscedastic (GARCH) \cite{bollerslev_generalized_1986} was introduced for efficient and accurate volatility prediction, researcher have also attempted to combine it with regime switching models in order to address structured market movements \citep{cai_markov_1994}. Previous literature has extensively studied market regimes, and proposed methods for modelling and forecasting. Markov decision processes have been proposed to model the transition probability between different market regimes, e.g. with the Hidden Markov Model (HMM) \citep{hamilton_new_1989}. Therefore, the Markov Regime-switching GARCH (MR-GARCH) has been extensively explored \cite{marcucci_forecasting_2005}. MR-GARCH leverages HMM to estimate two regimes with different levels of volatility, and makes forecasting individually on each regime. This convenient multi-period-ahead volatility recursive procedure forecasting has been proven to be significantly effective in out-of-sample prediction compared to single-regime GARCH, which is usually overly smooth and high in volatility forecasting.

However, this approach is limited by the curse of dimensionality, as the dimensionality of the hidden states is linear to the number of assets considered \citep{bekaert_how_2002}. On the other hand, some researchers believe that the market can be expressed as mixed multivariate distributions, and each state effectively corresponds to a distribution. Hence, temporal clustering methods such as Gaussian Mixture \citep{reynolds_gaussian_2015}, and K-Nearest Neighbors (KNN) \citep{kumar_comparative_2018} have been applied for this purpose. Yet, these methods are often based on strong assumptions and they are not originally designed for time-series problems, which results in both theoretical and practical issues. E.g., Gaussian Mixture assumes Gaussian nature in all base distributions, and KNN overlooks temporal consistency between single data points. An alternative are nonparametric kernel regression approaches as in e.g. \cite{franke_statistics_2019}. For estimating the IV surface, we will later rely on these techniques.

In addition, previous literature has started to look for alternative methods to cluster similar temporal data points into the same group based on certain comparison criteria. Subsequent clustering uses a sliding window to capture a period of data points and analyze it for recurrent patterns \citep{aghabozorgi_time-series_2015}. Alternatively, point clustering, instead of measuring spatial similarity between two slices of time-series, looks at each temporal point individually, and assigns this multivariate observation to an appropriate cluster based on a distance metrics \citep{kanungo_efficient_2002}. In 2017, \cite{hallac_toeplitz_2017} proposed the Toeplitz Inverse Covariance Clustering (TICC) algorithm, originally devised for electric vehicles action sensor. It classifies states based on the likelihood measures of short subsequences of observations and a corresponding sparse precision matrix under a Toeplitz constraint. Inspired by TICC, \cite{procacci_forecasting_2019} proposed a closely related methodology named Inverse Covariance Clustering (ICC), which is used as a main clustering methodology in this paper. This approach provides a point clustering of observations while enforcing temporal consistency by penalizing switching between states. The main advantages of ICC compared to TICC are its flexibility in the selection of similarity measures, and the use of sparse precision matrices via information filtering networks \citep{tumminello_tool_2005,massara_network_2017, massara_learning_2019,aste_sparse_2017} to reduce noise in financial time-series data. Hence, in this paper, we replace HNN with ICC as the main methodology for market regime clustering.

\section{Methodology} \label{sec:methodology}
The goal of this section is to describe how we use ICC to fit ISV models. The main difference between ISV and classical SV models is that we use some shape characteristics (moneyness and time-to-maturity) of the IV surface instead of historical return data. Nevertheless, there are jumps in returns and volatility that need to be accounted for. We will therefore use the clusters obtained by ICC and fit an ISVM for each cluster individually.

\subsection{Implied Stochastic Volatility Modeling}
The setup of the ISVM as introduced by Ait-Sahalia, Li, and Li in \cite{ait-sahalia_implied_2021} is a generic continuous bi-variate SV model with jumps. Under an assumed risk-neutral measure, we have $S_t$ the asset price and $v_t$ the corresponding volatility. They jointly follow a diffusion process
\begin{subequations}
\begin{align}
    &\frac{dS_t}{S_t-} = \left(r-d-\theta(v_t)\Bar{\mu}\right)dt+v_tdW_{1t}+\exp{\left(J_t-1\right)}dN_t, \label{eq:1a} \\
    &dv_t=\mu(v_t)dt+\gamma(v_t)dW_{1t}+\eta(v_t)W_{2t}. \label{eq:1b} 
\end{align}
\end{subequations}
The variables $r$ and $d$ denote the risk-free rate and the dividend yield of the underlying. They are assumed to be constant and observable. $W_{1t}$ and $W_{2t}$ are standard Brownian motions and independent of each other. $N_t$ is a doubly stochastic Poisson or Cox process with stochastic intensity $\theta(v_t)$. Denote $J_t$ as the size of the log-price jump. It is assumed to be independent of the asset price $S_t$. Once a jump occurs, $S_t$ changes according to $\log S_t - \log S_{t-} = J_t$, or $S_t - S_{t-} = (\exp{(J_t)}-1)S_{t-}$. $S_{t-}$ represents the asset price before a jump. 

In every step in time $l$ we observe $n_l$ implied volatilities $\Sigma^{data}$ as well as time-to-maturity $\tau_l^{(m)}$ and log-moneyness $k_l^{(m)}$ for $m=1,2,\dots,n_l$. Absent of jumps, the scalar functions $\mu$, $\gamma$, and $\eta$ can be fitted from the implied volatility surface through the bivariate regression:

\begin{align*}
    \Sigma^{data} (\tau_l^{(m)},k_l^{(m)}) &= \beta_l^{(1,0)} + \beta_l^{(0,0)}\tau_l^{(m)} + \beta_l^{(2,0)}(\tau_l^{(m)})^2 + \beta_l^{(0,1)}k_l^{(m)} \\
    & + \beta_l^{(1,1)}\tau_l^{(m)}k_l^{(m)}+ \beta_l^{(2,1)}(\tau_l^{(m)})^2k_l^{(m)} \\ 
    & + \beta_l^{(0,2)}(k_l^{(m)})^2 + \epsilon_l^{(m)}. \numberthis \label{eq:2} 
\end{align*}
However, if we want to additionally fit the jump terms there are some empirical challenges. It is practically impossible to observe third order characteristics of the IV surface, which are necessary for estimating the ISVM with jumps. We would require extremely high frequency data that is with the current liquidity of BTC options not available. Additionally, very short maturity options would have to be accurately observed. This is difficult due to the presence of negative powers that let the out-of-money IV possibly go to infinity as time-to-maturity shrinks to zero \citep{carr_what_2003}.

Since jumps effectively represent a regime change in the characteristics of the data that we seek to estimate, we propose a clustering algorithm as a pre-filtering mechanism to circumvent the empirical challenges that come with trying to estimate an SV model in the presence of jumps. 

\subsection{Inverse Covariance Clustering}

ICC is a penalized temporal clustering method, originated from the TICC algorithm proposed by \cite{hallac_toeplitz_2017}. This multivariate clustering method depends on the choice of a gain function, which incorporates the mean value of variables as well as the covariance matrix addressing the interdependent temporal co-variation. For a multivariate time-series of $n$ assets with $\mathbf r_t \in \mathbb R^{1\times n}$ the vector of returns at time $t$. The corresponding vector of their expected values can be expressed as $\boldsymbol \mu = \mbox{\sf E}(\mathbf r_t) \in \mathbb R^{1\times n}$ and accordingly, their covariance matrix is expressed $\boldsymbol{\Sigma} = \mbox{\sf E}\left[(\mathbf r_t-\boldsymbol \mu)^\top (\mathbf r_t-\boldsymbol \mu)\right]\in \mathbb R^{n\times n}$.

The gain function of the ICC is a measure that qualifies the gain when the time $t$ returns, $\mathbf r_t$, are associated with cluster $k$. ICC gathers in cluster $k$ observations that have the largest gain in such a cluster with respect to any other cluster: $G_{t,k} > G_{t,h}$ for all $h\not= k$. Two distance associated gain functions have been proposed in \cite{procacci_forecasting_2019}. $G^{Eu}_{t,k}$ is minus the square of the euclidean distance between the observation and the centroid of cluster $k$, which is expressed as:
\begin{equation}\label{eq:likelihodEuclidean}
    G^{Eu}_{t,k} =   -(\mathbf {r}_t-\widehat {\boldsymbol \mu}_k)(\mathbf {r}_t-\widehat {\boldsymbol \mu}_k)^\top
\end{equation}
where $\widehat {\boldsymbol \mu}_k$ is the sample mean return computed from the observations in cluster $k$. 
In this paper, we use a distance associated with the likelihood for multivariate normal distributions, which is instead
\begin{equation}\label{eq:likelihodN}
    G^{No}_{t,k} =  \frac{1}{2} \log | \widehat {\boldsymbol \Sigma}_k^{-1} | - n \frac{d^2_{t,k}}{2},
\end{equation}
with
\begin{equation}\label{eq:Maha}
d^2_{t,k} = (\mathbf{r}_t-\widehat {\boldsymbol \mu}_k)^\top \widehat {\boldsymbol \Sigma}_k^{-1}(\mathbf{r}_t-\widehat {\boldsymbol \mu}_k)
\end{equation}
the Mahalanobis distance where $\widehat {\boldsymbol \Sigma}_k$ is the sample covariance computed form the observations in cluster $k$.

The ICC approach uses the sparse inverse covariance that was shown to improve results considerably over the full covariance, as it is less sensitive to noise especially in financial time-series data. We use a TMFG information filtering graph proposed by \cite{massara_network_2017} for sparsification, where the local-global (LoGo) inversion procedure is described in \cite{barfuss_parsimonious_2016}. 

A final key element of the ICC methodology is the temporal consistency penalization of the cluster that is imposed by penalizing frequent switches between clusters. The assignment of the temporal instance $t$ to a cluster number, $k_t$, is performed iteratively starting from an initial random cluster assignment. Specifically we evaluate the penalized gain
\begin{equation}\label{eq:PenGain}
    \tilde G_{t,k_t} =  G_{t,k_t} - \lambda \delta_{k_{t-1},k_t},
\end{equation}
and assign observation $t$ to the cluster with largest penalized gain. The scalar $\lambda\in \mathbb R$ represents a penalty for switching between clusters. In the previous expression, $\delta_{k_{t-1},k_t}$ is the Kronecker delta returning one if $k_{t-1}=k_t$ and zero otherwise. After the assignment of the time-$t$ observation to a given cluster $k_t$, all cluster parameters (means and covariances) are recomputed with the new cluster assignments. 

\subsection{The MR-ISVM approach}
Recall that we would like to estimate the jump components in (\ref{eq:1a}), which is infeasible empirically. It is thus natural to try to estimate the parameters from a statistical approximation of the surface. Unfortunately, applying any regression to the full dataset does not result in realistic or accurate predictions due to the effect of jumps and nonstationarity. Jumps constitute temporary regime changes, which significantly alter the parameters we aim to estimate. We seek to overcome this problem by using the sketched clustering algorithm. 

The methodology hinges on the assumption that there is a finite number of regimes $R_1,\dots,R_k$ with correspondingly different regression parameters $(\beta^{(a,b)}{i,1},\dots,\beta^{(a,b)}{i,m})_{1 \leq i \leq k}$. Here, $\beta^{(a,b)}$ is the estimate of coefficient $\sigma^{(a,b)}$ with $(a,b)$ a nonnegative integer pair for the order of the respective coefficient. Further, we assume that our observation period can be subdivided into finite intervals in which the observed process stays in one regime. Phrased differently, there are time points $0 = t_0 < t_1 < \dots < t_n = T$ such that $S_t$ is in some fixed regime $R$ for all $t \in [t_{i},t_{i+1}]$. Now, given a temporally ordered data set $K$, we attempt to partition $$K = \bigsqcup K_i,$$ into a disjoint union of clusters $K_i = \{k(t_j) \in K \colon S(t_j) \in R_i\}$. 

Given an accurate such clustering, we can use the data points in each cluster $K_i$ to obtain a faithful estimate of the bivariate regression \ref{eq:2}. The fitted surfaces are then used to nonparametrically estimate the functions $\mu(v_t),$ $\gamma(v_t)$, and $\eta(v_t)$.
The chosen clustering method is particularly well-suited to the outlined task by both penalizing spatial and temporal distance. Ideally, for data of successively higher frequencies in the same time period, the number of clusters should stabilize. Frequent switches between clusters indicate either a highly volatile process, an unsuitable clustering method or that our assumptions regarding the observability of different regimes are erroneous. 

\section{Data} \label{sec:data}
We extract the most recent Deribit order book data from the BRC at the highest frequency of 20 minutes. The data ranges from April 13, 2021 to March 7, 2022 and consists of 9,116,870 observations from 12,274 assets. We conduct the experiment in two steps. First, we use ICC for obtaining clusters based on close to maturity options with a maturity of one week or less. Then, we fit an ISV model for each cluster separately on options with a maturity of 5 to 60 days. As outlined in the previous chapters, this approach resembles a SV model, however based on IV data, and with accounting for different IV regimes through employing ICC beforehand. An alternative approach would be using models that include jumps such as the SVCJ. A comparison between the two approaches at hand would be interesting. We leave this open for future research as it exceeds the scope of this publication.

Unlike previous approaches, we extend the lower maturity threshold from 15 to 5 days to account for the 24/7 trading of Bitcoin and adapt to our high frequency data set. The intuition behind choosing close to maturity for clustering the market is that these options reflect current beliefs about the market most accurately. Any options that are further from the money are only likely to be traded if market participants in the short term believe that Bitcoin is volatile enough for reaching distant price levels. In other words, we seek to take advantage of the fact that the knowledge of informed traders is priced into Deribit options, and that IV movements are an indicator of short-term movements in the market.

During clustering, we consider only those options that have a moneyness between 0.8 and 1.2 to exclude highly illiquid options, and filter out any options that have less than 66\% of data points, as we impute missing data points in the first step. The threshold can be altered if needed, however due to clustering based on the sparsified covariance matrix of implied volatilities, the imputation has little effect on the final result. By setting a threshold we therefore merely ensure that only liquid options are included in our dataset. 
$K$ and $\lambda$ strongly influence the fitted ISV models, as the number of clusters, and how stable we seek to keep them has a direct influence on each model's input data. In general, $K$ should be kept small for the sake of simplicity, but especially longer time series require a higher number of clusters. Assuming that different regimes exist, the choice is mostly between 2 and 3 empirically.
$\lambda$ can be treated as a typical tuning parameter such that it's size depends on the data and controls for the balance between stability and number of observations per cluster. To make sure that clustering can be achieved, we tested different values of $\lambda$ ranging from $-5$ to $5$ and ended up at value of $\lambda=0.5$. We initiate the procedure by verifying that at least one iteration $w$ can be performed with $\lambda=0$. This makes sure that the dataset actually has different clusters. In general, $\lambda$ should be kept as closely as possible to 0 if the data allows for it. When choosing large positive values of $\lambda$, the data seems to become more homogeneous in terms of cluster association, s.t. at some threshold value all data points will belong to a single cluster. When choosing very large negative values, the assignment to the different clusters seems indistinguishable from randomly assigning the points to the clusters. For each value of $\lambda \neq 0 $ in case no iteration was successful, we scale the threshold down to $75\%$ of its previous value until the data can be clustered. Thanks to this iterative procedure, we can start at a relatively large threshold and stop once convergence is achieved. This both saves computation time and prevents biased results.

We consider only options with a maturity between 5 and 60 days for the ISV model. As in \cite{ait-sahalia_implied_2021} we apply preprocessing to our data, such that we additionally restrict log-moneyness to be within $\pm v_t \sqrt{\tau}$, where, $\tau$ is the annualized time-to-maturity and $v_t$ is the instantaneous volatility. The instantaneous volatility is estimated by that observed IV where both time-to-maturity $\tau$ and log-moneyness $k$ are the closest to 0 at each observation. Additionally, to guarantee stability in the nonparametric regression, we only evaluate those time periods where each cluster has at least 25 observations. Eventually, we fit a total of 21,277 IV surfaces at a frequency of 20 minutes on 83 rolling window data sets in the observed time period.  We replicate the bootstrap estimator for the standard errors as in the original paper and generate 500 bootstrap samples, where we sample an IV surface for each day with the same number of IV observations as in the original data. 
\section{Results} \label{sec:results}

Remember that our goal is to fit an ISV model that helps us incorporate investor expectation through using IV data in the model. As an illustrative example, we randomly select the period of January 23 - 28, 2022 throughout this chapter. 
\begin{figure}[ht]
\centering
\begin{subfigure}{.3\textwidth}
\includegraphics[width=\textwidth]{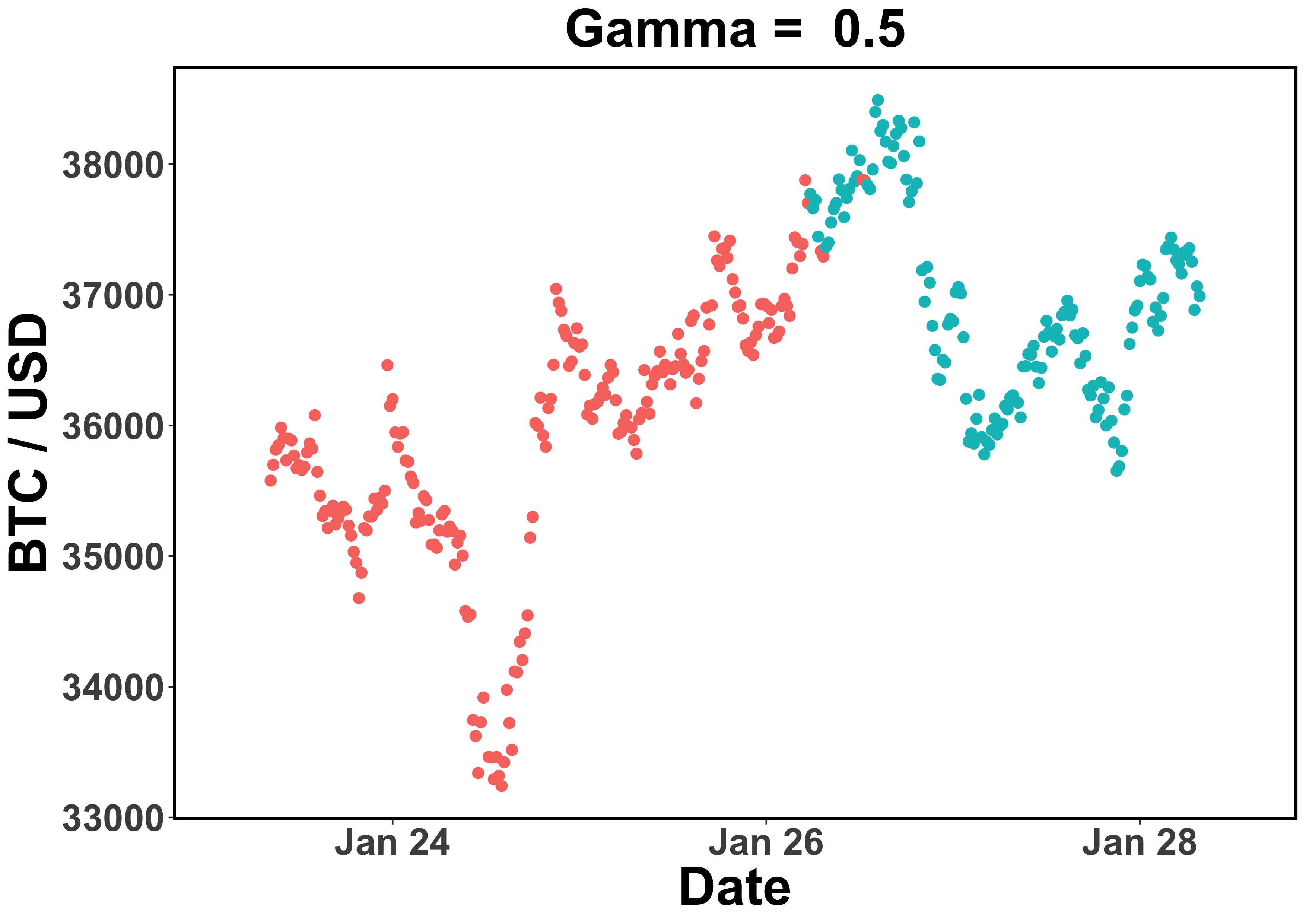} 
\includegraphics[width=\textwidth]{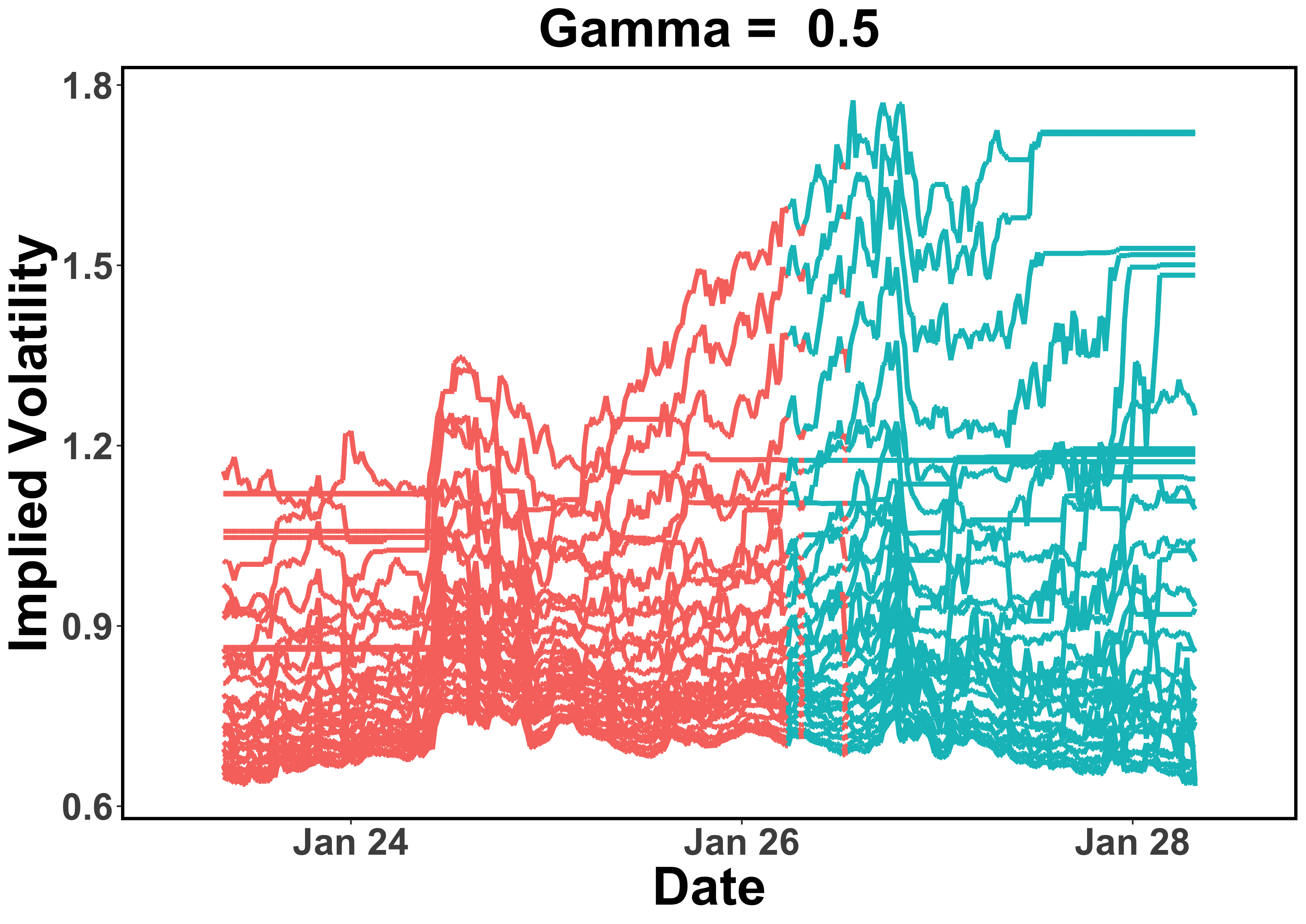}
\end{subfigure}
\caption{Bitcoin price and implied volatility with detected clusters. }
\label{fig:btc_cluster_both}
\end{figure}
Figure \ref{fig:btc_cluster_both} shows the identified clusters in the data. The upper plot shows the price evolution of the BTC/USD exchange rate during the observed time frame, and the lower plot shows the observed IV of all options considered for clustering. In the plot with $K=2$, the majority of observations belong to the first cluster (red) during a period of larger volatility and price movements, whereas the second cluster (green) contains those observations where the price and IV remain relatively stable. Next, we will investigate the effects of clustering on the ISVM fit.

\begin{figure}[htbp]
\centering
\begin{subfigure}{.45\textwidth}
\includegraphics[width=\textwidth]{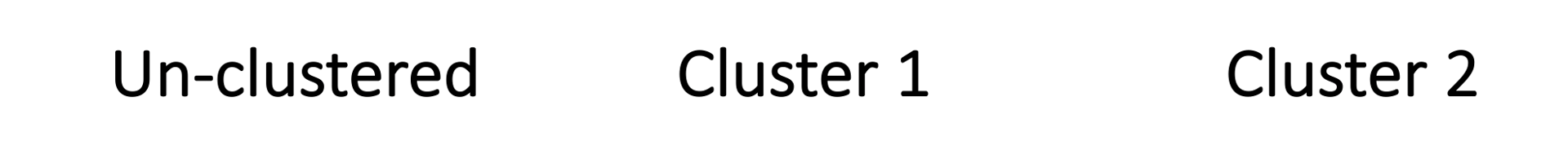} 
\end{subfigure}

\begin{subfigure}{.45\textwidth}
\includegraphics[width=\textwidth]{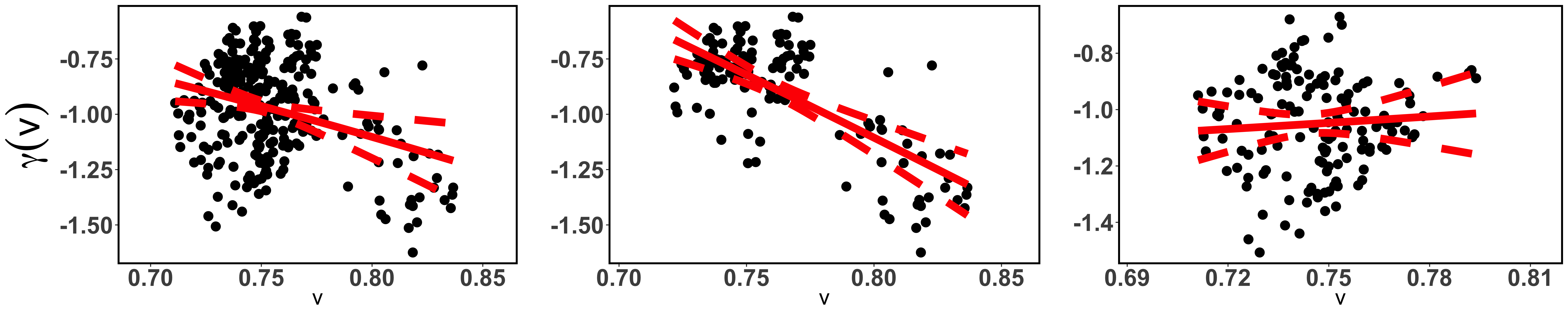} 
\caption{$\gamma(v)$}
\end{subfigure}

\begin{subfigure}{.45\textwidth}
\includegraphics[width=\textwidth]{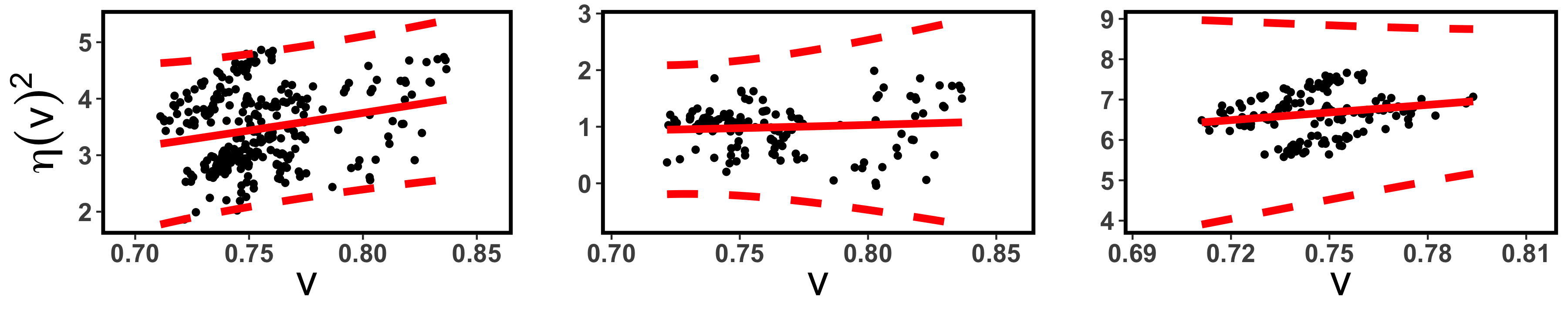}
\caption{$\eta(v)^2$}
\end{subfigure}

\begin{subfigure}{.45\textwidth}
\includegraphics[width=\textwidth]{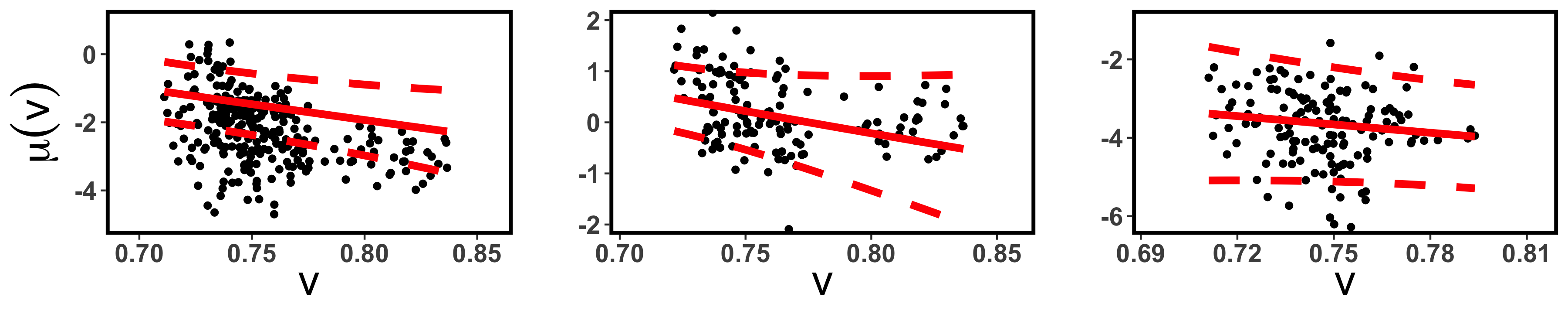}
\caption{$\mu(v)$}
\end{subfigure}

\caption{ISVM on data fitted on the whole dataset (first column) and on each cluster individually. }
\label{fig:btc_ISVM_clustered_K2}
\end{figure}

Figure \ref{fig:btc_ISVM_clustered_K2} shows a comparison of the nonparametric estimation of the functions $\mu$, $\gamma$, and $\eta^2$ from the data, based on the whole dataset against a fit with $K=2$. Every panel represents the fit of one of the functions. In each panel, the plot on the very left shows the fit on the whole data as a comparison, and the remaining plots show the fits for each cluster individually. The dots in each plot represent the data of the underlying function. The solid red lines are the mean estimates based on local regression and the dashed lines are the two standard deviation confidence intervals obtained from the Bootstrap. The model fit with clustering matches the data more closely due to less biased estimates and lower variance. Moreover, the model fits for $K_1$ and $K_2$ are distinct enough to conclude that at least some of the patterns within the data have been captured. Consider $\widehat{\mu}(.)$, which for whole dataset is negative. When $K=2$, in $K_1$, as long as the input argument is small the values remain positive. However, as the input argument grows larger the values become negative. In $K_2$, the direction is similar, but all values are negative. This indicates a different regime within the data that is consistent with the visual exploration of the clusters w.r.t. the data. $\widehat{\gamma}(.)$ is approximately linear, and the slope of the estimate differs between clusters. $\widehat{\eta^2}(.)$ is always positive and also approximately linear, however with very different levels across clusters. We conducted the same experiment with $K=3$. Since the conclusions are largely similar, we refrain from discussing that case further.

The effects we observe are largely similar to those in the original ISVM paper, especially when we introduce clustering, which indicates an overall well fitted model. MR-ISVM has allowed us to fit the IV surface of Bitcoin options during different volatility regimes. Consequently, we can use the stochastic volatility for pricing options based on the valuable information of informed traders that the implied volatility surface on Deribit contains.

Table \ref{tab:modelevalrmse} shows the root mean squared error (RMSE) of the nonparametric regression estimates of $\eta^2$, $\gamma$, and $\mu$ when $K=1$, and $\eta^2_1$ is the function estimated in cluster 1 when $K=2$, $\eta^2_2$ the function in cluster 2, etc. Note that while the table says $K=1$, this merely refers to the case where no clustering is performed and we will denote $K=1$ as the unclustered case from here on.  The first column shows the function, the second to fourth columns the mean, 5th, and the 95th percentile values. The final column shows the difference between the 95th and 5th percentile values as an indicator of how spread out the distribution of RMSE values is. For the unclustered case, we see that $\eta^2$ has the highest spread, and $\mu$ the lowest. 

The table shows that the clustering approach is effective in capturing different patterns in the data that are a result of nonstationarity, which is reflected by the heterogeneity of RMSE values across functions and between clusters. Even though in mean, all RMSE are lower than the benchmark values, there seems to be a difference in stability of RMSE between clusters. Consider $\eta^2_2$, the spread between percentiles is significantly lower than in $\eta^2_1$, and when $K=1$, while the spread for $\eta^2_1$ is approximately similar to the spread when $K=1$. The results for $\mu$ show a similar pattern when comparing $\mu_1$ (lower mean, lower spread) and $\mu_2$ (higher mean, higher spread). $gamma_1$ and $gamma_2$ seem to slightly deviate from this pattern. While the spread for $gamma_1$ is lower than the benchmark, the spread for $gamma_2$ is significantly higher. These results are not surprising as the basic idea of the MR-ISVM is to capture periods with different volatility regimes. In the case of $K=2$, it is apparent that there is always one regime with lower (higher) variation. Accordingly, the variation of RMSE values is also lower (higher). 

We also computed the mean absolute error (MAE) as an alternative metric, as well as the both error measures for $K=3$. For the sake of conciseness we do not discuss these results as the outcome was largely similar to the RMSE results with $K=2$. The results show that using clustering significantly reduces the estimation error. For one of the clusters the MR based method additionally reduces the variation of RMSE values. This demonstrates that the MR-ISVM approach is well suited for handling the nonstationarity that is prevalent in BTC option data. That is, without having to explicitly fit a jump component which is theoretically possible, but infeasible empirically, we can nevertheless fit models that account for nonstationarity and jumps in the data in an efficient way.

\begin{table}[ht]
\centering
\caption{Comparison of RMSE statistics with $K=1$ and $K=2$.} 
\label{tab:modelevalrmse}
\begin{tabular}{lrrrr}
  \hline
func & Mean & Pctile[5] & Pctile[95] & Diff Pctile[95]-[5] \\ 
  \hline
  $\eta^2$ & 0.49 & 0.18 & 1.23 & 1.05 \\ 
  $\eta^2_1$ & 0.42 & 0.16 & 1.14 & 0.98 \\ 
  $\eta^2_2$ & 0.37 & 0.13 & 0.93 & 0.80 \\ 
  \midrule
  $\gamma$ & 0.19 & 0.10 & 0.33 & 0.23 \\ 
  $\gamma_1$ & 0.16 & 0.09 & 0.29 & 0.20 \\ 
  $\gamma_2$ & 0.17 & 0.06 & 0.39 & 0.33 \\ 
  \midrule
  $\mu$ & 0.66 & 0.30 & 1.22 & 0.92 \\ 
  $\mu_1$ & 0.57 & 0.28 & 1.02 & 0.74 \\ 
  $\mu_2$ & 0.58 & 0.18 & 1.13 & 0.95 \\ 
  \hline
\end{tabular}
\end{table}

\FloatBarrier 

\section{Conclusion} \label{sec:conclusion}
We proposed the MR-ISVM approach and discussed in details an example of Bitcoin options with different volatility regimes. This work addressed three major gaps in the literature. First, cryptocurrencies are highly sentiment-driven which makes the usage of historical data for option pricing models inaccurate. We circumvented this by using implied volatility instead to include investor expectations and accounting for the buying pressure caused by informed traders. Second, the underlying data is nonstationary and subject to frequent jumps, which makes any model without jumps inaccurate. The identification of different regimes through clustering has the advantage that we do not need to explicitly account for the unique market microstructure of cryptocurrencies when accounting for jumps in prices or volatility, or make complex model extensions to overcome the empirical and theoretical challenges in including a jump component. Third, we add more interpretability to the ICC approach by exploring the effect on a series of nonparametrically estimated functions. The interpretation of the cluster outcomes has proven to be difficult when the data is complex or the number of clusters is high.

Overall, we have demonstrated that by using temporal clustering can successfully separate the data into different volatility regimes, where the fitting of pricing model has significant lower estimation errors than fitting with the entire period. This effect is illustrated by the consistently reduced mean estimation errors in the clusters. Moreover, the significantly reduced variation in the estimation errors suggests that clustering also helps to stabilize the previously inaccurate regression estimates that could not capture jumps. Therefore, the proposed MR-ISVM approach allows us to fit option pricing models based on the ISVM when the data contains jumps, even when jumps can not be explicitly addressed in the underlying model.

While we have shown an example of how these methods can be combined, future work could more systematically examine the effect of $K$, the number of cluster, and $\lambda$, the temporal consistency penalty, for maximizing the model fit. Future work is also planned to benchmark other commonly used model in the literature and industry, e.g., SVCJ.

\newpage

\bibliographystyle{ACM-Reference-Format}
\bibliography{BTCVola}
\end{document}